\documentclass[a4paper,fleqn,usenatbib]{mnras}

\usepackage{newtxtext,newtxmath}
\usepackage[T1]{fontenc}
\usepackage{ae,aecompl}

\usepackage{graphicx}	% Including figure files
\usepackage{amsmath}	% Advanced maths commands
\usepackage{amssymb}	% Extra maths symbols
\usepackage{pdflscape}

\title[High-velocity gas in W28]{High-velocity interstellar absorption associated with the supernova remnant W28}

\author[A. M. Ritchey]{
Adam M. Ritchey,$^{1,2}$\thanks{E-mail: aritchey@astro.washington.edu}
\\
$^{1}$Department of Astronomy, University of Washington, Seattle, WA 98195\\
$^{2}$Eureka Scientific, 2452 Delmer, Suite 100, Oakland, CA 96402
}

\date{Accepted XXX. Received YYY; in original form ZZZ}

\pubyear{2020}

\begin{document}
\label{firstpage}
\pagerange{\pageref{firstpage}--\pageref{lastpage}}
\maketitle

\begin{abstract}
We present an analysis of moderately high resolution optical spectra obtained for the sight line to CD$-$23~13777, an O9 supergiant that probes high velocity interstellar gas associated with the supernova remnant W28. Absorption components at both high positive and high negative velocity are seen in the interstellar Na~{\sc i}~D and Ca~{\sc ii}~H and K lines toward CD$-$23~13777. The high velocity components exhibit low Na~{\sc i}/Ca~{\sc ii} ratios, suggesting efficient grain destruction by shock sputtering. High column densities of CH$^+$, and high CH$^+$/CH ratios, for the components seen at lower velocity may be indicative of enhanced turbulence in the clouds interacting with W28. The highest positive and negative velocities of the components seen in Na~{\sc i} and Ca~{\sc ii} absorption toward CD$-$23~13777 imply that the velocity of the blast wave associated with W28 is at least 150~km~s$^{-1}$, a value that is significantly higher than most previous estimates. The line of sight to CD$-$23~13777 passes very close to a well-known site of interaction between the SNR and a molecular cloud to the northeast. The northeast molecular cloud exhibits broad molecular line emission, OH maser emission from numerous locations, and bright extended GeV and TeV $\gamma$-ray emission. The sight line to CD$-$23~13777 is thus a unique and valuable probe of the interaction between W28 and dense molecular gas in its environs. Future observations at UV and visible wavelengths will help to better constrain the abundances, kinematics, and physical conditions in the shocked and quiescent gas along this line of sight.
\end{abstract}

\begin{keywords}
ISM: abundances -- ISM: atoms -- ISM: molecules -- ISM: kinematics and dynamics -- ISM: supernova remnants
\end{keywords}

%%%%%%%%%%%%%%%%%%%%%%%%%%%%%%%%%%%%%%%%%%%%%%%%%%

%%%%%%%%%%%%%%%%% BODY OF PAPER %%%%%%%%%%%%%%%%%%

\section{Introduction}\label{sec:introduction}
The remnants of supernova explosions are shaped by the interstellar environments in which the explosions occur. While the majority of supernova remnants (SNRs) are classified as shell-type (Green 2019), meaning that their X-ray and radio morphologies have a shell-like appearance, a significant fraction of remnants belong to a class of ``mixed-morphology'' SNRs (Rho \& Petre 1998). These remnants, also known as ``thermal composites'' (Jones et al.~1998), have shell-like radio morphologies but their X-ray emission is centrally concentrated. In contrast to other types of composite remnants where the center-filled X-ray emission is nonthermal in origin and powered by a central pulsar, the X-ray emission from mixed-morphology SNRs is primarily thermal. Various models, invoking phenomena such as cloud evaporation (White \& Long 1991) and thermal conduction (Cox et al.~1999; Shelton et al.~1999), have been proposed to explain the origin of the hot interiors of mixed morphology SNRs. A common characteristic of this class of remnants is that they are observed to be interacting with molecular clouds. The center-filled X-ray morphologies are likely a consequence of the interaction with dense gas. However, the precise mechanism, or combination of mechanisms, by which these remnants attain their distinct morphologies is not yet understood.

The SNR W28 is an archetype of the class of mixed morphology SNRs. A distinct radio shell is seen predominantly to the north and east of the remnant (e.g., Dubner et al.~2000), while thermal X-ray emission is concentrated toward the center (e.g., Rho \& Borkowski 2002). A secondary X-ray peak is seen toward the northeast near a well-known site of interaction between the SNR and a molecular cloud (see, e.g., Nicholas et al.~2012). A fainter X-ray shell component is observed toward the southwest. Arikawa et al.~(1999) mapped the northeast molecular cloud in CO (3$-$2) and (1$-$0) emission. By separating narrow and broad emission components, they demonstrated the positional offset between shocked and unshocked molecular gas. The northeast cloud seems to have been partially overtaken by the supernova blast wave. A secondary ridge of shocked molecular material lies to the north (e.g., Arikawa et al.~1999; Nicholas et al.~2011). Enhanced velocity dispersions on the side of the northeast cloud facing the center of W28 are clear indications of external disruption of the cloud by the SNR (Nicholas et al.~2011, 2012; Maxted et al.~2016). Forty-one 1720 MHz OH masers are found to lie along the edges of the shocked molecular material (Claussen et al.~1997). The maser emission is yet another clear indication of the interaction between the SN shock and dense molecular gas associated with W28.

Aharonian et al.~(2008) discovered very high energy (VHE) $\gamma$-ray emission coincident with molecular clouds in the W28 region. Using the High Energy Stereoscopic System (HESS), Aharonian et al.~(2008) identified four bright TeV $\gamma$-ray sources near W28. One source (labelled HESS J1801$-$233) coincides with the shocked northeast molecular cloud. The other three sources (labelled HESS J1800$-$240A, B, and C) are located to the south of W28 beyond the outer boundary of the radio shell which delineates the present location of the SN shock front (see Figure~\ref{fig:w28_targets}). An extended GeV $\gamma$-ray source coincides with the northern HESS source and with the shocked molecular cloud (Abdo et al.~2010; Giuliani et al.~2010). Fainter GeV $\gamma$-ray emission is found to be associated with the southern HESS sources (Abdo et al.~2010; Giuliani et al.~2010; Hanabata et al.~2014; Cui et al.~2018). The $\gamma$-ray emission likely results from the decay of neutral pions produced through interactions between shock-accelerated cosmic rays and dense molecular gas. The southern HESS sources have been interpreted as ambient molecular clouds illuminated by cosmic rays that escaped from the shell of W28 when the remnant was much younger (Hanabata et al.~2014; Cui et al.~2018). It is not yet clear if the $\gamma$-ray emission from the northeast shocked cloud is also due to escaped cosmic rays or to newly-accelerated cosmic rays that are now interacting with dense gas in the post-shock region.

Numerous investigations have been carried out to examine the structure and kinematics of the shocked molecular clumps associated with W28 (e.g., DeNoyer 1983; Arikawa et al.~1999; Reach \& Rho 2000; Reach et al.~2005; Nicholas et al.~2011, 2012; Gusdorf et al.~2012; Maxted et al.~2016). These studies provide information on the characteristics of the shocks (i.e., shock type, shock velocity, pre-shock density) driven into the dense clumps by the SN blast wave. Considerably less information is available on the velocity of the SN blast wave itself. Lozinskaya (1974) detected an expansion velocity of 40--50~km~s$^{-1}$ from observations of H$\alpha$ and [N~{\sc ii}] emission lines in W28. The radial velocities of individual [N~{\sc ii}] filaments are as high as 60 to 80~km~s$^{-1}$ (Lozinskaya 1980). Shock velocities between 60 and 90~km~s$^{-1}$ were inferred by Bohigas et al.~(1983) from optical line ratios in comparison with shock models. From their own observations of optical emission lines, Long et al.~(1991) found that the shock velocity should be larger than $\sim$70~km~s$^{-1}$. A shell of neutral hydrogen surrounding W28 was found by Vel\'{a}zquez et al.~(2002) to have an expansion velocity of $\sim$30~km~s$^{-1}$. The H~{\sc i} shell was interpreted as swept-up interstellar material and thus might be expected to have a velocity significantly less than that of the SN blast wave.

While the optical observations of W28 suggest a blast wave velocity of $\sim$100~km~s$^{-1}$, much larger values are implied by the temperatures of the northeast and southwest X-ray shell components. Rho \& Borkowski (2002), for example, obtained X-ray temperatures of $\sim$0.6~keV and $\sim$1.2--1.5~keV for the northeast and southwest shells, respectively. If this X-ray emission results from shocked-heated gas in a region immediately behind the SN blast wave, then the implied shock velocities would range from 700 to 1100~km~s$^{-1}$. Rho \& Borkowski (2002) interpreted the X-ray emission as ``fossil radiation'' from a time when the remnant was younger and the shock velocity was higher, attributing the properties of the radiation to some combination of cloud evaporation, thermal conduction, and mixing due to various hydrodynamic instabilities. Still, the high X-ray temperatures observed in the interiors of mixed-morphology SNRs such as W28 remain poorly understood.

The distance to W28 has also been somewhat controversial. In part, this has been caused by disagreements concerning the systemic velocity of the clouds associated with W28. Arikawa et al.~(1999) detected two narrow CO emission components toward W28 with local standard of rest (LSR) velocities of +7 and +21~km~s$^{-1}$. The OH masers found to be associated with W28 exhibit centroid velocities in the range +5 to +15~km~s$^{-1}$ (Claussen et al.~1997). Lozinskaya (1974) reported a mean LSR velocity for the remnant as a whole of +18~km~s$^{-1}$, which corresponds to a kinematic distance of 3.6~kpc. However, the H~{\sc i} observations reported by Vel\'{a}zquez et al.~(2002) show a very strong self-absorption feature at +7~km~s$^{-1}$, which the authors identify as the atomic counterpart to the molecular cloud known to be interacting with W28. Adopting a systemic velocity of +7~km~s$^{-1}$, Vel\'{a}zquez et al.~(2002) derive a kinematic distance to W28 of $1.9\pm0.3\,{\rm kpc}$.

\begin{figure}
\includegraphics[width=\columnwidth]{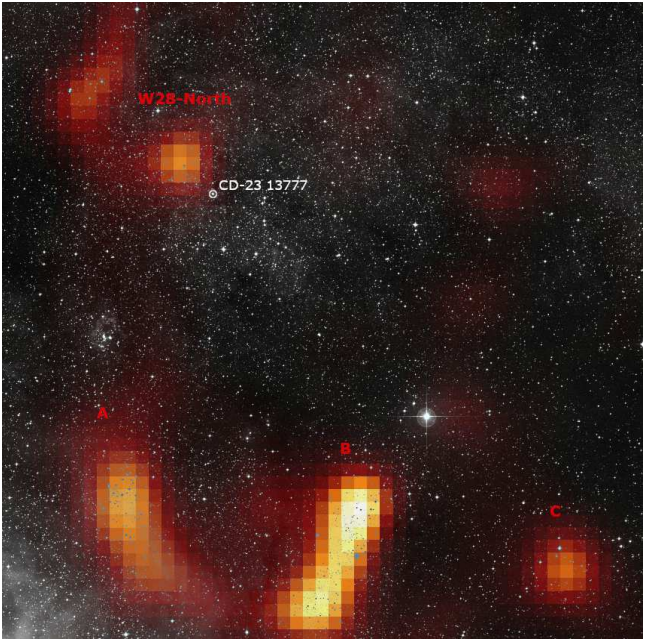}
\caption{Composite image of W28. The optical image from the Digitized Sky Survey (POSS-II/red filter) is shown superimposed onto the VHE $\gamma$-ray excess counts map obtained with HESS (Aharonian et al.~2008). The location of the background star CD$-$23~13777, which is the subject of the present investigation, is indicated. The TeV sources HESS J1801$-$233 (W28-North) and HESS J1800$-$240A, B, and C (A, B, and C) are also labelled.}
\label{fig:dss_hess}
\end{figure}

One approach to studying the kinematics and physical conditions in shocked and quiescent gas associated with SNRs is to examine the interstellar absorption features that appear in the ultraviolet and visible spectra of stars located behind the remnant. Such an approach has recently been used to obtain a detailed set of physical conditions in gas interacting with the SNR IC~443 (Ritchey et al.~2020). To date, there have been no published studies of interstellar absorption lines (in the UV or visible) toward stars in the W28 region. The purpose of the present investigation is to identify sight lines that can serve as probes of the interstellar material interacting with W28. To this end, we conducted a moderate-resolution ground-based spectroscopic survey of 15 stars in the W28 region. One star in particular (CD$-$23~13777) proved to be a remarkable probe of shock-accelerated gas in W28. Thus, the majority of this paper will focus on just this sight line. The on-sky location of CD$-$23~13777 in relation to the TeV $\gamma$-ray sources observed with HESS is shown in Figure~\ref{fig:dss_hess}. The remainder of this paper is organized as follows. The observations and procedures used for data reduction are discussed in Section~\ref{sec:observations}. Our basic analysis of the interstellar absorption lines detected toward CD$-$23~13777 is described in Section~\ref{subsec:col_den}. In Section~\ref{subsec:temp_var}, we present evidence of temporal variations in the Na~{\sc i} absorption components detected at high velocity toward CD$-$23~13777. Our results for this sight line are discussed in Section~\ref{sec:discussion} in the context of multiwavelength observations of the gas interacting with W28. Our basic conclusions are summarized in Section~\ref{sec:conclusions}. A more complete description of our spectroscopic survey of the W28 region is presented in Appendix~\ref{sec:appendix_a}. Detailed component structures for the interstellar species observed toward CD$-$23~13777 are tabulated in Appendix~\ref{sec:appendix_b}.

\section{Observations and Data Reduction}\label{sec:observations}
Observations of the star CD$-$23~13777 were originally obtained in 2011 as part of a survey of stars probing W28 using the Astrophysical Research Consortium echelle spectrograph (ARCES; Wang et al.~2003) on the 3.5~m telescope at Apache Point Observatory (APO). The echelle spectrograph provides complete wavelength coverage in the range 3800--10200~\AA{} at a resolving power of $R\approx31,500$ ($\Delta v\approx9.5\,{\rm km~s}^{-1}$). Fifteen early-type stars in the vicinity of W28 were observed in an effort to discern high-velocity interstellar absorption features that would indicate that the sight line to the star probed shocked gas associated with the SNR. However, only CD$-$23~13777 showed evidence of high-velocity absorption. Absorption components at velocities exceeding $100\,{\rm km~s}^{-1}$ are seen in the Na~{\sc i}~D lines and in the Ca~{\sc ii}~H and K lines toward CD$-$23~13777. In most of the other directions surveyed, the Na~{\sc i} and Ca~{\sc ii} absorption components exhibit only very modest velocities, with typical values being between $-30$ and $+30\,{\rm km~s}^{-1}$. (The Na~{\sc i} absorption profiles toward all 15 observed stars are presented in Figure~\ref{fig:w28_spectra}.)

At the time of our original survey of the W28 region, the distances to the program stars were uncertain. Spectroscopic parallaxes indicated that the survey stars were at least as far away as the remnant at 1.9~kpc (Vel\'{a}zquez et al.~2002). However, now that trigonometric parallaxes are available for these stars from the \emph{Gaia} satellite (e.g., Bailer-Jones et al.~2018), their distances are much better constrained. The \emph{Gaia} Data Release 2 (DR2) parallaxes reveal that all but two of the stars surveyed have distances of less than 1.8~kpc (see Table~\ref{tab:arces_targets}), placing them in the foreground of the remnant. The most distant star that we observed is CD$-$23~13777, whose \emph{Gaia} DR2 parallax indicates a distance of $2.4\pm0.3\,{\rm kpc}$. The discovery of high-velocity gas toward CD$-$23~13777, but not toward the other observed stars, helps to confirm that the distance to W28 is $\sim$2~kpc.

Having identified CD$-$23~13777 as a unique probe of high-velocity gas associated with W28, we sought to improve the signal-to-noise (S/N) ratio of the data by acquiring additional ARCES spectra of this star. Ultimately, we obtained 13 separate 30 minute exposures of CD$-$23~13777 over the course of five nights between 2011 September and 2015 August. Individual exposure times were limited to 30 minutes to minimize the number of cosmic-ray hits occurring during each integration. The raw data were reduced using {\sc iraf} following standard procedures for ARCES spectra (see, e.g., Ritchey \& Wallerstein 2012). Telluric absorption lines were removed from the calibrated spectra using observations of the unreddened standard star $\eta$~UMa, which was observed on each of the five nights when spectra of CD$-$23~13777 were obtained. The corrected spectra were shifted to the LSR frame of reference and the individual exposures were coadded to produce a final high S/N ratio spectrum. In order to search for temporal variations in the interstellar absorption profiles, we also produced nightly sum spectra by coadding the two or three exposures obtained on a given night (see Section~\ref{subsec:temp_var}). Total exposure times and S/N ratios achieved for each of the five nights are listed in Table~\ref{tab:observations}.

\begin{table}
\centering
\caption{Log of APO observations of the star CD$-$23~13777. For each set of observations, we give the UT date, the total exposure time (in seconds), and the S/N ratio per (2.5 pixel) resolution element near the Na~{\sc i}~D lines.}
\label{tab:observations}
\begin{tabular}{lcc}
\hline
UT Date & Exp.~Time & S/N \\
 & (s) & \\
\hline
2011 Sept 7 & 3600 & 470 \\
2012 Aug 26 & 3600 & 330 \\
2013 Apr 24 & 5400 & 300 \\
2015 May 24 & 5400 & 550 \\
2015 Aug 20 & 5400 & 410 \\
\hline
\end{tabular}
\end{table}

\section{Analysis}\label{sec:analysis}
\subsection{Column Densities and Component Structure}\label{subsec:col_den}
\begin{figure}
\includegraphics[width=\columnwidth]{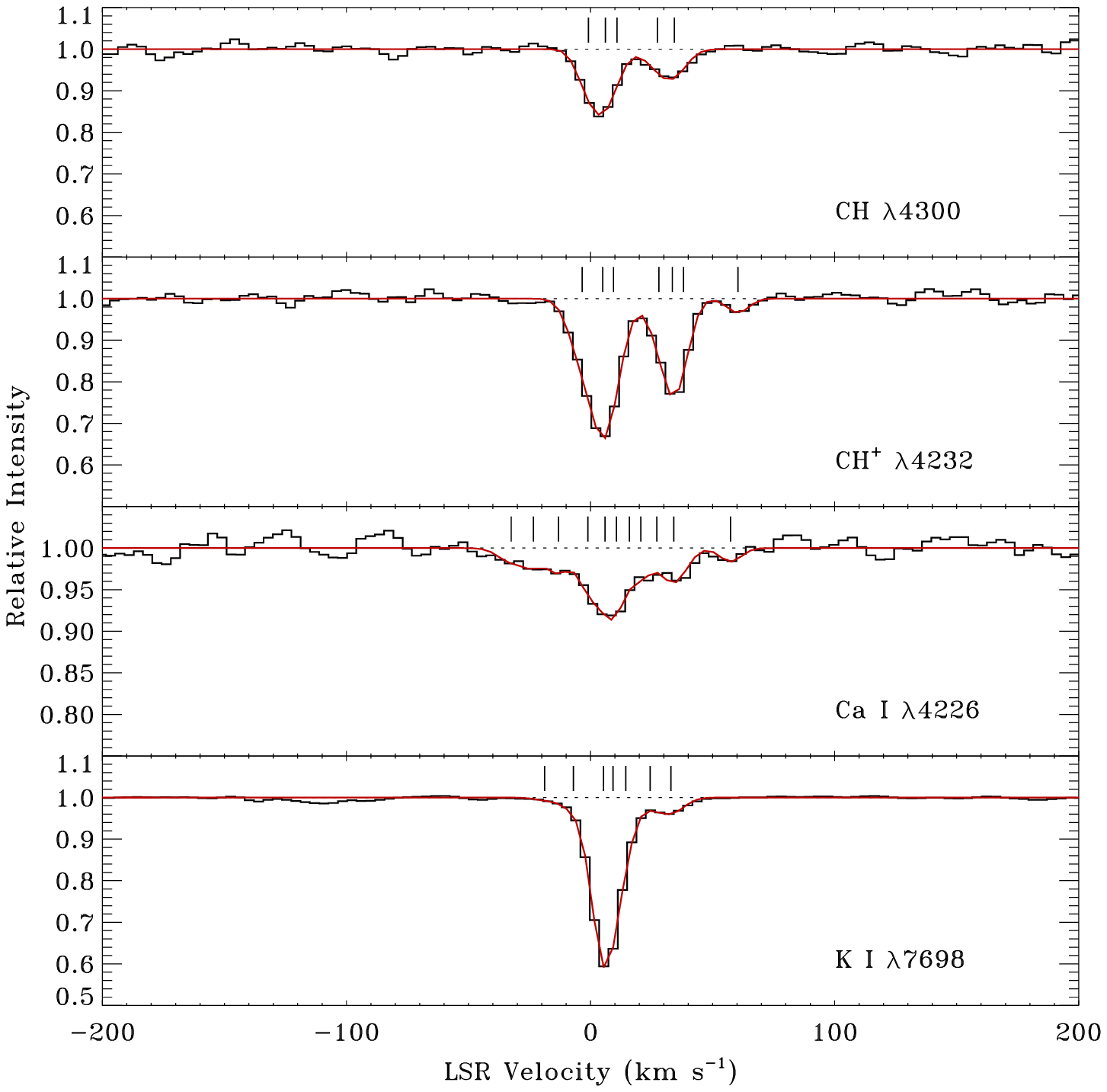}
\caption{Profile synthesis fits to the CH~$\lambda4300$, CH$^+$~$\lambda4232$, Ca~{\sc i}~$\lambda4226$, and K~{\sc i}~$\lambda7698$ lines toward CD$-$23~13777. Synthetic absorption profiles (red curves) are shown superimposed onto the observed spectra (black histograms). Tick marks give the positions of the individual components included in the fits.}
\label{fig:profile_fits1}
\end{figure}

\begin{figure}
\includegraphics[width=\columnwidth]{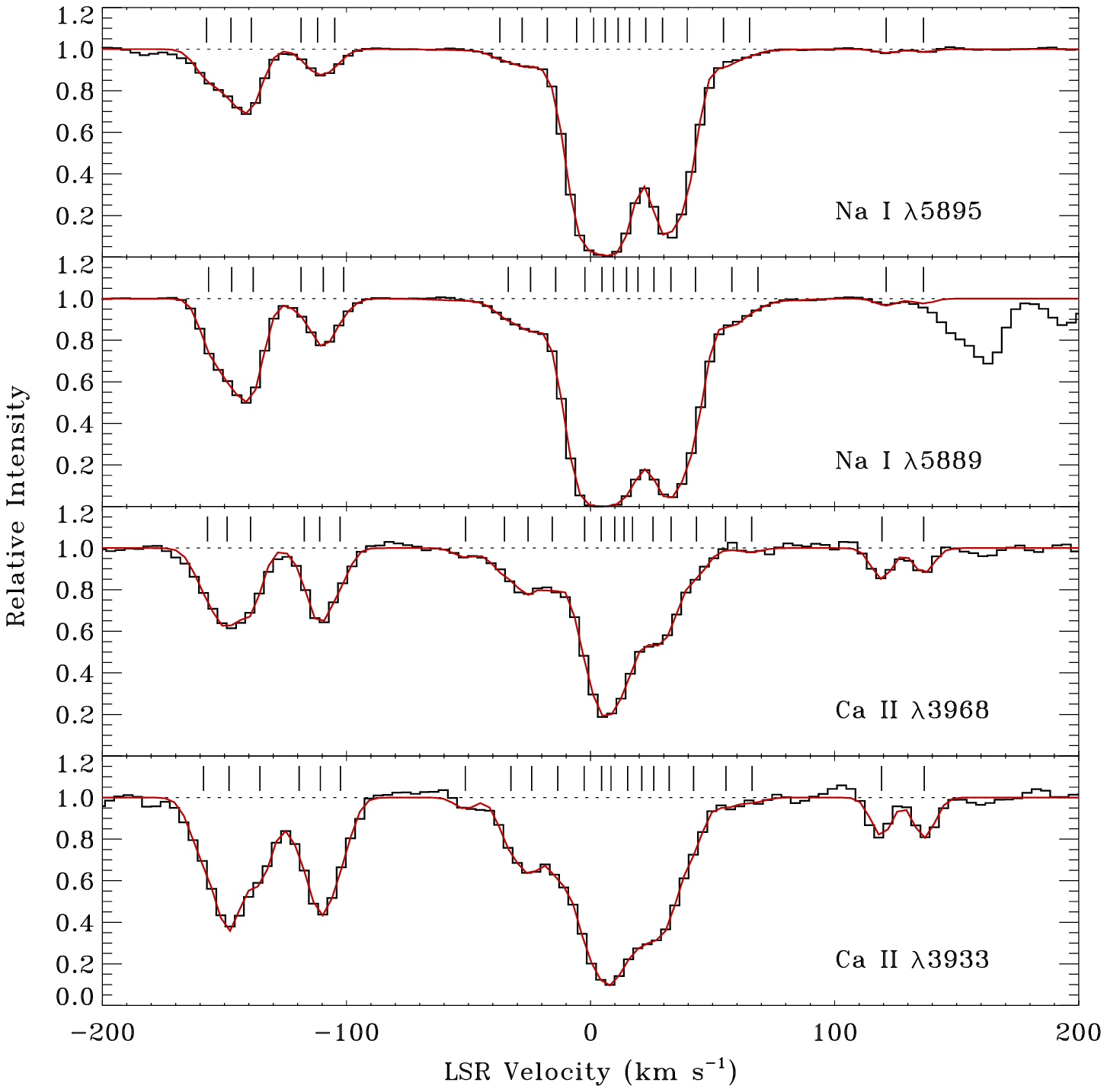}
\caption{Profile synthesis fits to the Na~{\sc i}~$\lambda\lambda5889,5895$ and Ca~{\sc ii}~$\lambda\lambda3933,3968$ lines toward CD$-$23~13777. Synthetic absorption profiles (red curves) are shown superimposed onto the observed spectra (black histograms). Tick marks give the positions of the individual components included in the fits.}
\label{fig:profile_fits2}
\end{figure}

The most notable features in the ARCES spectra of CD$-$23~13777 (see Figures~\ref{fig:profile_fits1} and \ref{fig:profile_fits2}) are the high-velocity components detected in the interstellar Na~{\sc i} and Ca~{\sc ii} absorption profiles. Two relatively strong absorption features are seen at LSR velocities of $-$146 and $-$110~km~s$^{-1}$, while two weaker components are found at velocities of +120 and +137~km~s$^{-1}$. As would be expected, the dominant absorption complex is found near 0~km~s$^{-1}$. We do not detect any of the high-velocity features in the absorption profiles of the trace neutral species K~{\sc i} or Ca~{\sc i}, nor do we see these components in the absorption profiles of the molecular species CH or CH$^+$. All four of these latter species, however, show absorption at more moderate velocities. We do not find any convincing evidence for absorption from the CN $B$$-$$X$ (0$-$0) band near 3874~\AA{} nor from the $A$$-$$X$ (2$-$0) band of C$_2$ near 8757~\AA{}.

Small portions of the coadded spectrum surrounding the detected interstellar lines were isolated and continuum normalized via low-order polynomial fits to the line-free regions. The absorption profiles were then analyzed by means of the multi-component profile synthesis routine {\sc ismod} (see Sheffer et al.~2008), which assumes a Voigt profile shape for each component included in the fit. The profile synthesis routine returns the best-fitting column density, Doppler $b$-value, and velocity of each component through a simple root-mean-square minimizing technique. However, ARCES spectra present a particular challenge when analyzing interstellar lines due to the coarse resolution of the data. High and ultra-high resolution observations of the interstellar medium (ISM) have revealed that individual interstellar components are typically characterized by very narrow intrinsic widths (i.e., $b\lesssim2\,{\rm km~s}^{-1}$ for trace neutral species such as Na~{\sc i} and K~{\sc i}; Welty \& Hobbs 2001; Price et al.~2001; Crawford 2001). Such narrow components will not be resolved by ARCES, which has a velocity resolution of $\sim$9.5~km~s$^{-1}$. Thus, when fitting the absorption profiles of the lines observed toward CD$-$23~13777, we have included a sufficient number of components so that the derived $b$-values fall within a range typical of interstellar lines (i.e., $0.5\gtrsim b \gtrsim4.5\,{\rm km~s}^{-1}$). In most cases, the lines are not severely saturated so that the adopted component structure will not have too large an effect on the column densities obtained.

The results of our profile synthesis fits for the prominent interstellar species detected toward CD$-$23~13777 are presented in Figures~\ref{fig:profile_fits1} and \ref{fig:profile_fits2}. (The component parameters derived from these fits are tabulated in Appendix~\ref{sec:appendix_b}.) In fitting the main Na~{\sc i} absorption complexes at low velocity (which are severely saturated), we used the component structure found from the K~{\sc i}~$\lambda7698$ line as a guide. However, while the resulting $N$(Na~{\sc i})/$N$(K~{\sc i}) ratios appear to be consistent with the typical interstellar value of $\sim$90 (e.g., Welty \& Hobbs 2001), we still consider the Na~{\sc i} column densities associated with these low velocity components to be highly uncertain. Another complication that arose in fitting the Na~{\sc i} profiles is that the weak components at high positive velocity in the $\lambda5889$ line are partially blended with the stronger components at high negative velocity in the $\lambda5895$ line (see Figure~\ref{fig:profile_fits2}). We therefore used the high positive velocity features seen in the $\lambda5895$ line as templates for fitting and removing the corresponding components in the $\lambda5889$ profile.

\begin{figure}
\includegraphics[width=\columnwidth]{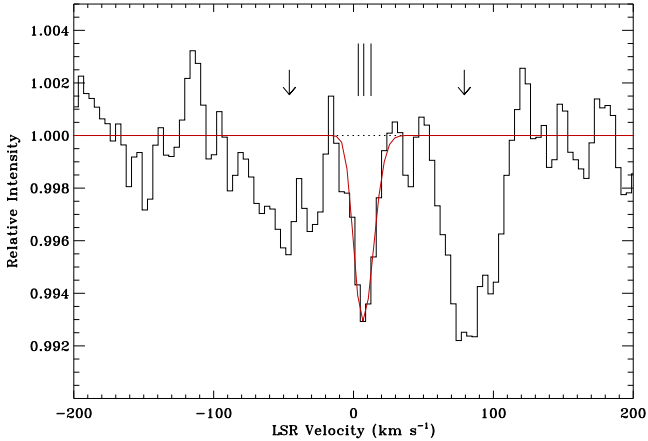}
\caption{Profile synthesis fit to the Li~{\sc i}~$\lambda6707$ line toward CD$-$23~13777. A template for the Li~{\sc i} line was created based on the three strongest components observed in K~{\sc i}. Arrows mark the positions of two weak DIBs that have (rest-frame) wavelengths of 6706.6 and 6709.4~\AA{}.}
\label{fig:li_fit}
\end{figure}

Since the sight line to CD$-$23~13777 is adjacent to what is likely a site of interaction between shock-accelerated cosmic rays and molecular gas (Figure~\ref{fig:dss_hess}), and since Li is a by-product of such interactions (e.g., Meneguzzi et al.~1971), it is important to establish whether Li~{\sc i} absorption is detectable in this direction. In Figure~\ref{fig:li_fit}, we show the region of the coadded spectrum of CD$-$23~13777 surrounding the Li~{\sc i}~$\lambda6707$ transition. The very weak Li~{\sc i} feature is detected at a significance level of greater than $5\sigma$. However, due to the coarse resolution of the ARCES spectra, we are unable to resolve the individual fine-structure lines of the Li~{\sc i} doublet, which have a velocity separation of only 6.7~km~s$^{-1}$. Nor can we evaluate the Li isotope ratio from the ARCES data. We can, however, determine the total Li~{\sc i} column density along the line of sight. To accomplish this, we created a template for the Li~{\sc i} absorption profile based on the three strongest components seen in the K~{\sc i} line (see Table~\ref{tab:comp_struct}). We then fit this template to the observed Li~{\sc i} spectrum, keeping the relative velocities, fractional column densities, and $b$-values of the components fixed. The result of this fit is shown by the solid red line in Figure~\ref{fig:li_fit}. The figure also shows two weak diffuse interstellar bands (DIBs) that are detected on either side of the Li~{\sc i} feature. (Many other DIBs are detected along the line of sight to CD$-$23~13777. However, the analysis of these features is beyond the scope of the present paper.)

The total equivalent widths and line-of-sight column densities of the atomic and molecular species observed toward CD$-$23~13777 are presented in Table~\ref{tab:totals}. (We do not list column densities for the Na~{\sc i}~D lines because the line profiles are severely saturated at low velocity rendering the total column densities uncertain.) The equivalent width errors reported in Table~\ref{tab:totals} account for both photon noise and errors in continuum placement, while the column density errors include an additional term based on the degree of saturation in the line profile. We note that there may be some unresolved saturated absorption that is not fully accounted for in our fits to the Ca~{\sc ii} profiles since the stronger line of the Ca ~{\sc ii} doublet yields a somewhat smaller column density than the weaker line. However, the two results for the total Ca~{\sc ii} column density agree with one another at about the $2\sigma$ level.

\begin{table}
\centering
\caption{Total equivalent widths (in m\AA{}) and column densities (in cm$^{-2}$) of the atomic and molecular species observed toward CD$-$23~13777. (No column densities are reported for the Na~{\sc i}~D lines due to significant saturation in the line profiles at low velocity.)}
\label{tab:totals}
\begin{tabular}{lccc}
\hline
Species & $\lambda$ & $W_{\lambda}$ & $\log N$ \\
 & (\AA{}) & (m\AA{}) & \\
\hline
K~{\sc i} & 7698.965 & $182.8\pm1.6$ & $12.14\pm0.02$ \\
Li~{\sc i} & 6707.826 & \phantom{11}$2.8\pm0.5$ & \phantom{1}$9.97\pm0.07$ \\
Na~{\sc i} & 5889.951 & $1486.7\pm3.4$\phantom{1} & [sat.] \\
 & 5895.924 & $1205.4\pm3.4$\phantom{1} & [sat.] \\
Ca~{\sc i} & 4226.728 & \phantom{1}$47.6\pm4.9$ & $11.25\pm0.04$ \\
Ca~{\sc ii} & 3933.661 & $1034.1\pm12.8$ & $13.35\pm0.03$ \\
 & 3968.467 & \phantom{1}$692.5\pm10.5$ & $13.42\pm0.03$ \\
CH & 4300.313 & \phantom{1}$54.0\pm3.4$ & $13.85\pm0.03$ \\
CH$^+$ & 4232.548 & $139.5\pm4.1$ & $14.29\pm0.02$ \\
\hline
\end{tabular}
\end{table}

\begin{table*}
\centering
\caption{Column densities (in cm$^{-2}$) for distinct velocity components observed toward CD$-$23~13777. The velocities listed are the column-density weighted mean velocities (in km~s$^{-1}$) averaged over all of the species in which a given component is detected. The Ca~{\sc ii} and Na~{\sc i} column densities shown here were obtained by taking the weighted means of the results from the two lines of the doublets. (No column densities are given for the main Na~{\sc i} absorption complexes at +5 and +32~km~s$^{-1}$ because these features are too strongly saturated to allow for accurate column density determinations.)}
\label{tab:components}
\begin{tabular}{ccccccc}
\hline
$\langle v_{\rm LSR} \rangle$ & $\log N$(Ca~{\sc ii}) & $\log N$(Na~{\sc i}) & $\log N$(Ca~{\sc i}) & $\log N$(K~{\sc i}) & $\log N$(CH$^+$) & $\log N$(CH) \\
\hline
$-$146.0 & $12.59\pm0.02$ & $12.21\pm0.01$ & \ldots & \ldots & \ldots & \ldots \\
$-$109.9 & $12.43\pm0.03$ & $11.64\pm0.01$ & \ldots & \ldots & \ldots & \ldots \\
\phantom{1}$-$51.3 & $11.12\pm0.10$ & \ldots & \ldots & \ldots & \ldots & \ldots \\
\phantom{1}$-$28.7 & $12.13\pm0.03$ & $11.36\pm0.02$ & $10.35\pm0.14$ & \ldots & \ldots & \ldots \\
\phantom{11}+4.9 & $13.04\pm0.05$ & [sat.] & $11.07\pm0.05$ & $12.11\pm0.02$ & $14.08\pm0.02$ & $13.69\pm0.03$ \\
\phantom{1}+32.2 & $12.51\pm0.05$ & [sat.] & $10.46\pm0.11$ & $11.01\pm0.03$ & $13.85\pm0.03$ & $13.33\pm0.06$ \\
\phantom{1}+59.2 & $11.09\pm0.14$ & $11.29\pm0.07$ & \phantom{1}$9.96\pm0.20$ & \ldots & $12.81\pm0.11$ & \ldots \\
+120.2 & $11.78\pm0.05$ & $10.62\pm0.03$ & \ldots & \ldots & \ldots & \ldots \\
+136.5 & $11.69\pm0.11$ & $10.44\pm0.05$ & \ldots & \ldots & \ldots & \ldots \\
\hline
\end{tabular}
\end{table*}

In order to place our determinations of column densities for the line of sight to CD$-$23~13777 in the context of more general surveys of the ISM, we can compare our determinations with the trends revealed by those more general surveys. For example, the $N$(Li~{\sc i})/$N$(K~{\sc i}) ratio seen toward CD$-$23~13777 is fully consistent with the trend between these two species obtained from surveys of diffuse molecular clouds (e.g., Welty \& Hobbs 2001; Knauth et al.~2003). The total CH column density also seems to be consistent with the amount of K~{\sc i} present (e.g., Welty \& Hobbs 2001; Welty et al.~2006). On the other hand, the CH$^+$ column density toward CD$-$23~13777, which we find to be $N({\rm CH}^+)\approx2.0\times10^{14}\,{\rm cm}^{-2}$, is among the highest values known for sight lines probing diffuse clouds in the local Galactic ISM (e.g., Welty et al.~2006; Smoker et al.~2014). Still, while the $N$(CH$^+$)/$N$(CH) ratio, which equals $\sim$3 for the entire line of sight, is higher than average, it is not as high as observed toward some Galactic regions such as the Pleiades, where this ratio can reach values of 20 or more (Ritchey et al.~2006). The column density of Ca~{\sc i} also seems to be somewhat high compared to the amount of K~{\sc i} seen in this direction, but the $N$(Ca~{\sc i})/$N$(Ca~{\sc ii}) ratio is not unusual (e.g., Welty et al.~2003).

Further insight may be gained from an examination of the column densities in distinct velocity components along the line of sight. For this exercise, we will sum the column densities of any closely-spaced components that together constitute a single distinct feature. In this way, uncertainties in the detailed underlying component structure of the various species will not significantly affect our interpretation. In Table~\ref{tab:components}, we present the column densities of nine such distinct components identified toward CD$-$23~13777. The velocities listed in Table~\ref{tab:components} are the column-density weighted mean velocities averaged over all of the species in which a given component is detected (see Table~\ref{tab:comp_struct}). The Ca~{\sc ii} and Na~{\sc i} column densities were obtained by taking the weighted means of the results from the two lines of the doublets. (Again, we do not list column densities for the main Na~{\sc i} absorption complexes at +5 and +32~km~s$^{-1}$ due to significant saturation in the line profiles at these velocities.)

The $N$(Na~{\sc i})/$N$(Ca~{\sc ii}) ratio varies considerably among the various components identified toward CD$-$23~13777. The lowest values of this ratio are found for the high positive velocity components at +120 and +137~km~s$^{-1}$, where $N$(Na~{\sc i})/$N$(Ca~{\sc ii})~$\approx$~0.06. Somewhat higher values are found for the high negative velocity components. We find $N$(Na~{\sc i})/$N$(Ca~{\sc ii}) ratios of 0.42 and 0.16 for the components at $-$146 and $-$110~km~s$^{-1}$, respectively. The K~{\sc i} column densities derived for the main low velocity absorption complexes at +5 and +32~km~s$^{-1}$ imply respective $N$(Na~{\sc i})/$N$(Ca~{\sc ii}) ratios of $\sim$11 and $\sim$3 (assuming the ratio of Na~{\sc i} to K~{\sc i} is $\sim$90; e.g., Welty \& Hobbs 2001). Large variations in the $N$(Na~{\sc i})/$N$(Ca~{\sc ii}) ratio are typical along sight lines probing gas associated with SNRs. Similar variations have been seen, for example, toward stars behind the Vela SNR (Danks \& Sembach 1995), the Monoceros Loop SNR (Welsh et al.~2001), and the SNRs S147 (Sallmen \& Welsh 2004) and IC~443 (Welsh \& Sallmen 2003; Ritchey et al.~2020). Since Ca is usually heavily depleted onto interstellar dust grains (e.g., Crinklaw et al.~1994), while Na is only lightly depleted, the $N$(Na~{\sc i})/$N$(Ca~{\sc ii}) ratio can reach values as high as 10--100 in cold, quiescent interstellar clouds (e.g., Crawford 1992). The low $N$(Na~{\sc i})/$N$(Ca~{\sc ii}) ratios that are typical of high velocity gas associated with SNRs may then be ascribed to the destruction of dust grains by SN shocks, although ionization effects may also be important considering the difference in ionization potential between Na~{\sc i} and Ca~{\sc ii}.

The $N$(Ca~{\sc i})/$N$(Ca~{\sc ii}) ratio can be useful for placing constraints on the ionization balance in diffuse clouds independent of any depletion effects. However, for the line of sight to CD$-$23~13777, absorption from Ca~{\sc i} is detected only at low to moderate velocity. The $N$(Ca~{\sc i})/$N$(Ca~{\sc ii}) ratios that we find for the components at $-$29, +5, and +32~km~s$^{-1}$ are all $\sim$0.01, which is a typical value for sight lines probing diffuse clouds (e.g., Welty et al.~2003). The $N$(Ca~{\sc i})/$N$(Ca~{\sc ii}) ratio seems to be somewhat higher for the component at +59~km~s$^{-1}$. However, this determination is uncertain because the associated absorption features are very weak. It is unusual to see absorption from species such as Ca~{\sc i} and CH$^+$ at moderately high velocity (such as we see near +59~km~s$^{-1}$ toward CD$-$23~13777). However, similar moderately high velocity Ca~{\sc i} and CH$^+$ absorption components were detected along sight lines probing IC~443 (Hirschauer et al.~2009; Ritchey et al.~2020). The $N$(CH$^+$)/$N$(CH) ratios for both of the main low velocity components toward CD$-$23~13777 are significantly higher than the typical interstellar value of $\sim$0.9 (e.g., Welty et al.~2006). We find respective $N$(CH$^+$)/$N$(CH) ratios of $2.5\pm0.2$ and $3.3\pm0.5$ for the components at +5 and +32~km~s$^{-1}$. Detailed chemical models have revealed that the CH$^+$ abundance is a sensitive tracer of the dissipation of turbulence in diffuse clouds (e.g., Godard et al.~2009, 2014). The high $N$(CH$^+$)/$N$(CH) ratios we find for the low velocity components toward CD$-$23~13777 may therefore be indicative of enhanced turbulence in the gas associated with these absorption features.

\subsection{Temporal Variations}\label{subsec:temp_var}
As discussed in Section~\ref{sec:observations}, the ARCES spectra of CD$-$23~13777 were acquired on five separate occasions over the course of approximately four years. While this was done mainly in an effort to build up the S/N ratio in the final coadded spectrum, it also affords us the opportunity to determine whether any changes occurred in the interstellar absorption profiles during that four year period. To examine this issue, we produced separate coadded spectra of CD$-$23~13777 for each of the five nights when data were obtained. In searching for temporal changes, we will focus our effort on the Na~{\sc i}~D and Ca~{\sc ii}~H and K lines since these lines are intrinsically strong and are the only lines that show high velocity interstellar absorption. From an examination of the equivalent widths of the Ca~{\sc ii} absorption features in the nightly sum spectra, we find that any variations that may be present do not significantly rise above the level of the uncertainties. (In other words, any apparent changes in the Ca~{\sc ii} equivalent widths are at the level of $\sim$$3\sigma$ or less.) This is not entirely unexpected since the S/N ratios in the nightly coadded spectra are rather low in the vicinity of the Ca~{\sc ii} lines. (A typical value is $\sim$30.)

\begin{figure}
\includegraphics[width=\columnwidth]{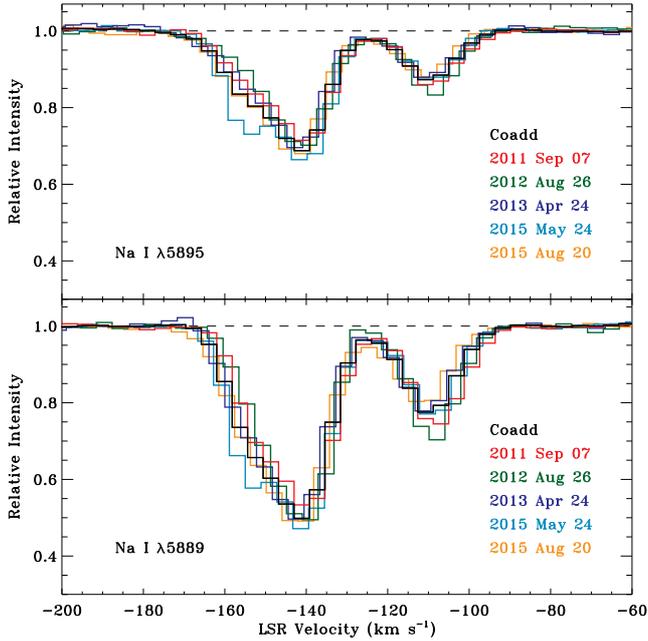}
\caption{Comparison of the Na~{\sc i} spectra taken on five separate nights between 2011 September and 2015 August for the high negative velocity absorption complex observed toward CD$-$23~13777. The final co-added spectrum is also shown. In these data, the high positive velocity components of the $\lambda5889$ line, which are partially blended with the high negative velocity features in the $\lambda5895$ profile, have been removed. A weakly variable component is evident in the spectrum taken on 2015 May 24.}
\label{fig:na_evol}
\end{figure}

Much higher S/N ratios are achieved in the nightly sum spectra of CD$-$23~13777 near the Na~{\sc i}~D lines (see Table~\ref{tab:observations}), meaning that any temporal changes will be easier to discern in Na~{\sc i}. While we do not find any convincing evidence of changes in the Na~{\sc i} absorption profiles for components at low to moderate velocity (i.e., $-30\lesssim v_{\rm LSR} \lesssim+60\,{\rm km~s}^{-1}$), we do see evidence for weak variations in the absorption components at high negative velocity. (The Na~{\sc i} components at high positive velocity are too weak to allow for any meaningful examination of temporal changes.) In Figure~\ref{fig:na_evol}, we plot the absorption profiles of the Na~{\sc i} complexes near $-$146 and $-$110~km~s$^{-1}$ as seen in the nightly sum spectra and in the final coadded spectrum. (In these data, the high positive velocity components of the $\lambda5889$ line, which are partially blended with the high negative velocity features in the $\lambda5895$ profile, have been removed.) Immediately apparent in the figure is the increase in absorption that occurred near $-$154~km~s$^{-1}$ in the spectrum taken on 2015 May 24. (It is important to note that the increase in absorption occurs in both of the Na~{\sc i}~D lines as would be expected if this is a real change and not just an artifact of noise or a telluric feature.) Indeed, it was the appearance of this weakly variable component in the May 24 spectrum that prompted us to acquire an additional spectrum of CD$-$23~13777 later that same year. In the subsequent spectrum taken on 2015 August 20, the Na~{\sc i} profiles near $-$154~km~s$^{-1}$ seem to have returned to their previous norm. However, upon further examination, other weak yet systematic changes can be seen in the Na~{\sc i} absorption profiles for the components found at high negative velocity. For example, absorption from the weaker complex at $-$110~km~s$^{-1}$ seems to vary systematically from a maximum on 2012 August 26 to a minimum on 2015 August 20.

In order to place more quantitative constraints on these observed changes, we analyzed the high negative velocity Na~{\sc i} absorption profiles from the nightly sum spectra by means of our profile fitting routine, adopting the same procedures as were used in fitting the coadded data. The resulting column densities of the two main absorption complexes seen at high negative velocity are listed in Table~\ref{tab:na_evol} for each of the five nights. As anticipated, the column density of the absorption complex near $-$146~km~s$^{-1}$ increased substantially in the spectrum taken on 2015 May 24 but then reverted to a more typical value by the time the next spectrum was obtained on 2015 August 20. Meanwhile, the column density of the absorption complex near $-$110~km~s$^{-1}$ reached a maximum in the spectrum taken on 2012 August 26 and then decreased considerably by 2013 April 24, maintaining this lower value over the next few years. (Again, these changes are seen in both lines of the Na~{\sc i} doublet simultaneously, which helps to confirm that these are real changes in column density rather than random variations due to noise or telluric interference.) The column-density weighted mean velocities of the two main absorption complexes seen at high negative velocity show only very minor changes over the course of the observations. Given the low resolution of the ARCES data, it is difficult to determine whether these changes in velocity are real. However, if the underlying component structure of a given absorption complex changes with time (for example, if one subcomponent becomes stronger relative to another), then small changes in the column-density weighted mean velocity would be expected.

Temporal changes in absorption components detected along sight lines probing SNRs are fairly common. The most famous examples are the many sight lines probing the Vela SNR that have shown dramatic changes in their Na~{\sc i} and Ca~{\sc ii} absorption profiles over time (e.g., Hobbs et al.~1991; Danks \& Sembach 1995; Cha \& Sembach 2000; Welty et al.~2008; Rao et al.~2016, 2017). Moderately high velocity time-variable Na~{\sc i} components have also recently been found toward stars probing the Monoceros Loop SNR (Dirks \& Meyer 2016) and IC~443 (Ritchey et al.~2020). While the variations we see in the high negative velocity Na~{\sc i} absorption components toward CD$-$23~13777 are perhaps not as dramatic as in some of these other examples, they nevertheless help to strengthen the connection between time-variable interstellar components and gas associated with SNRs.

\begin{table}
\centering
\caption{Column densities (in cm$^{-2}$) and column-density weighted mean velocities (in km~s$^{-1}$) for the Na~{\sc i} absorption complexes near $-$146 and $-$110~km~s$^{-1}$ as measured on five separate nights between 2011 September and 2015 August.}
\label{tab:na_evol}
\begin{tabular}{lcccc}
\hline
UT Date & $\langle v_{\rm LSR} \rangle$ & $\log N$ & $\langle v_{\rm LSR} \rangle$ & $\log N$ \\
\hline
2011 Sept 7 & $-$143.9 & $12.17\pm0.01$ & $-$108.6 & $11.71\pm0.02$ \\
2012 Aug 26 & $-$142.5 & $12.20\pm0.02$ & $-$108.6 & $11.77\pm0.03$ \\
2013 Apr 24 & $-$145.0 & $12.20\pm0.04$ & $-$110.1 & $11.62\pm0.02$ \\
2015 May 24 & $-$145.5 & $12.33\pm0.02$ & $-$109.4 & $11.64\pm0.02$ \\
2015 Aug 20 & $-$145.4 & $12.24\pm0.01$ & $-$112.0 & $11.60\pm0.02$ \\
\hline
\end{tabular}
\end{table}

\section{Discussion}\label{sec:discussion}
The most important result of our spectroscopic survey of stars in the W28 region is our discovery of high velocity interstellar absorption along the line of sight to CD$-$23~13777. Our detection of Na~{\sc i} and Ca~{\sc ii} absorption at both high positive and high negative velocity in this direction suggests that we are viewing both sides of the expanding shell of the remnant. The velocity of the dominant interstellar absorption component toward CD$-$23~13777 at +5~km~s$^{-1}$ is consistent (given the relatively low resolution of the ARCES spectra) with a systemic velocity for W28 of +7~km~s$^{-1}$ (Arikawa et al.~1999; Vel\'{a}zquez et al.~2002). Relative to the systemic velocity (assuming $v_{\rm sys}=+7\,{\rm km~s}^{-1}$) the highest positive and negative velocity absorption components toward CD$-$23~13777 have velocities of +130 and $-$153~km~s$^{-1}$ (see Table~\ref{tab:components}). This potentially suggests an asymmetry in the expansion of the SNR shell, with the shell expanding more rapidly on the near side than on the far side. The other major absorption components observed toward CD$-$23~13777 are distributed more symmetrically about the systemic velocity. The second highest positive and negative velocity components have velocities of +113 and $-$117~km~s$^{-1}$ relative to $v_{\rm sys}$. At the edges of the main absorption complex toward CD$-$23~13777 there is a pair of weak components with relative velocities of +52 and $-$58~km~s$^{-1}$. Adjacent to the main interstellar component is another pair of components with relative velocities of +25 and $-$36~km~s$^{-1}$. This last pair exhibits velocities that are similar to the expansion velocity inferred for the H~{\sc i} shell observed by Vel\'{a}zquez et al.~(2002).

The velocities of the high velocity components discovered toward CD$-$23~13777 are significantly higher than most previous estimates of the velocity of the shock associated with W28. Observations of optical emission lines in W28 have generally suggested shock velocities of $v_s\lesssim100\,{\rm km~s}^{-1}$ (Lozinskaya 1974; Bohigas et al.~1983; Long et al.~1991). Our observations indicate that the shock velocity could be 150~km~s$^{-1}$ or greater. Note that the velocities observed along the line of sight to CD$-$23~13777 are simply the radial velocities, which represent lower limits to the actual shock velocities if there are significant transverse components to the motion. Furthermore, the gas motions we are detecting in absorption may be due to shocks driven into pre-existing interstellar clouds by the SN blast wave. In such a scenario, the cloud shocks will generally have velocities that are less than that of the blast wave moving through the lower density intercloud medium. The blast wave velocity might therefore be considerably higher than 150~km~s$^{-1}$.

The position of the sight line to the background star CD$-$23~13777 seems to have been particularly fortuitous. This sight line is located very close to the edge of the shocked northeast molecular cloud on the side of the cloud facing the interior of the SNR (see Figure~\ref{fig:w28_targets}). The velocity dispersion distributions measured for various molecular species associated with the northeast cloud (Nicholas et al.~2011, 2012; Maxted et al.~2016) strongly suggest that the interior face of the cloud was the point of contact between the cloud and the SN shock. At the interface between the shocked and unshocked portions of the northeast cloud, a half ring of 1720 MHz OH masers is found with velocities in the range +7 to +14~km~s$^{-1}$ (Claussen et al.~1997; see also Nicholas et al.~2012). These velocities are similar to those of the main interstellar components toward CD$-$23~13777 (see Table~\ref{tab:comp_struct}). The northeast molecular cloud is also a site of strong GeV and TeV $\gamma$-ray emission (Aharonian et al.~2008; Abdo et al.~2010; Giuliani et al.~2010) and therefore most likely a site of cosmic-ray acceleration. The shocked portion of the northeast cloud might thus be expected to exhibit an enhanced Li abundance since Li is produced, in part, by spallation and fusion reactions between cosmic ray particles and interstellar nuclei (Meneguzzi et al.~1971; Lemoine et al.~1998). While the Li~{\sc i}/K~{\sc i} ratio toward CD$-$23~13777 does not appear to be unusual, a better probe of Li production is the $^7$Li/$^6$Li isotope ratio, which we are unable to determine due to the low resolution of the ARCES spectra.

Additional observations of CD$-$23~13777 are needed to discern the physical conditions in the high velocity absorption components detected in this direction. Spectroscopic observations in the UV, using \emph{HST}/COS for example, would allow the densities, temperatures, and thermal pressures of the components to be determined through use of the many UV diagnostics available in that wavelength regime. An analysis of this kind, using UV observations of atomic fine-structure lines to discern the physical conditions in shocked and quiescent gas, has recently been applied to sight lines in IC~443 (Ritchey et al.~2020). In that study, the high velocity components were found to exhibit a combination of enhanced thermal pressures and significantly reduced dust-grain depletions, indicating that the absorption is tracing gas in a cooling region far downstream from shocks driven into neutral gas clumps. A similar scenario may explain the high velocity components observed toward CD$-$23~13777. However, the near symmetry of the components seen at high positive and high negative velocity could indicate that a dense shell is forming as the remnant enters the radiative phase of SNR evolution (e.g., Chevalier 1999).

Ritchey et al.~(2020) reported the detection of a very high velocity absorption component in highly-ionized species toward a star behind IC~443. The velocity of this component (which is $\sim$600~km~s$^{-1}$) is consistent with the shock velocities inferred from observations of soft thermal X-ray emission from IC~443 (e.g., Troja et al.~2006), but is much higher than the shock velocity of $\sim$100~km~s$^{-1}$ typically quoted for this remnant. IC~443 is a mixed-morphology SNR like W28. Thus, it would be interesting to determine whether sight lines through W28 (such as that toward CD$-$23~13777) also exhibit absorption from highly-ionized species at velocities close to those derived from the temperatures of the X-ray shell components (e.g., Rho \& Borkowski 2002). Such a determination would require absorption-line observations in the UV.

Higher resolution ground-based spectra of CD$-$23~13777 near the Na~{\sc i} and Ca~{\sc ii} lines would enable the complex component structure of those absorption features to be examined in much greater detail than is possible from the ARCES spectra. Moreover, spectroscopic monitoring of the absorption complex at high negative velocity would help to confirm the temporal variations seen in the ARCES data. High-resolution spectra at very high S/N would be required to determine the Li isotope ratio in the main interstellar absorption component toward CD$-$23~13777. A lower than average $^7$Li/$^6$Li ratio, as was detected for a line of sight in IC~443 (Taylor et al.~2012), would be expected in the case of Li production by cosmic rays. A measurement of the Li isotope ratio toward CD$-$23~13777 could help to clarify the relationship between the +7~km~s$^{-1}$ molecular cloud in W28 and the source of the cosmic rays responsible for the $\gamma$-ray emission.

\section{Conclusions}\label{sec:conclusions}
Moderately high resolution optical spectra of 15 early-type stars in the vicinity of W28 were acquired to search for high-velocity interstellar absorption components associated with shocked gas in the SNR. Along one line of sight, toward the background star CD$-$23~13777, we find numerous Na~{\sc i} and Ca~{\sc ii} absorption components at high positive and high negative velocity. The Na~{\sc i}/Ca~{\sc ii} ratios in these high velocity components are significantly lower than those determined for the components at low velocity, suggesting that the high velocity components have been subjected to grain destruction by shock sputtering. The high column densities of CH$^+$, and the high CH$^+$/CH ratios, for the components found at relatively low velocity toward CD$-$23~13777 may be indicative of enhanced turbulence in the clouds interacting with W28. The highest positive and negative velocities of the Na~{\sc i} and Ca~{\sc ii} components detected toward CD$-$23~13777 imply that the velocity of the shock wave associated with W28 is at least 150~km~s$^{-1}$. This is a significantly higher value for the shock velocity than most previous observations of W28 have indicated. The detection of high-velocity interstellar absorption toward CD$-$23~13777, but not toward the other observed stars, helps to confirm that the distance to W28 is $\sim$2~kpc. Temporal changes in the Na~{\sc i} absorption complex seen at high negative velocity toward CD$-$23~13777 serve to strengthen the connection between time-variable interstellar components and gas associated with SNRs.

The line of sight to CD$-$23~13777 passes very close to the apparent point of contact between the SNR shock and the northeast molecular cloud in W28. This cloud exhibits broad molecular line emission, OH maser emission from numerous locations, and bright extended GeV and TeV $\gamma$-ray emission. The CD$-$23~13777 sight line is thus a unique and valuable probe of the interaction between W28 and dense molecular gas in its environment. Future observations of CD$-$23~13777 at UV wavelengths will enable a more detailed examination of the physical conditions in the shocked and quiescent components along the line of sight. In addition, high-resolution ground-based spectra of CD$-$23~13777 acquired at very high S/N will allow a determination of the Li isotope ratio, which could provide insight into the diffusion of cosmic rays accelerated at the SN shock and their subsequent interaction with dense molecular gas in the vicinity of W28.

\section*{Acknowledgements}
We thank Steve Federman for providing useful feedback on an early draft of this paper. Support for this research was provided by the Kenilworth Fund of the New York Community Trust. The results presented here are based on observations obtained with the Apache Point Observatory 3.5 m telescope, which is owned and operated by the Astrophysical Research Consortium. We acknowledge use of the SIMBAD database operated at CDS, France.

%%%%%%%%%%%%%%%%%%%%%%%%%%%%%%%%%%%%%%%%%%%%%%%%%%



%\bibliographystyle{mnras}


%%%%%%%%%%%%%%%%%%%%%%%%%%%%%%%%%%%%%%%%%%%%%%%%%%

%%%%%%%%%%%%%%%%% APPENDICES %%%%%%%%%%%%%%%%%%%%%

\appendix

\section{ARCES survey of the W28 region}\label{sec:appendix_a}
\begin{figure}
\includegraphics[width=\columnwidth]{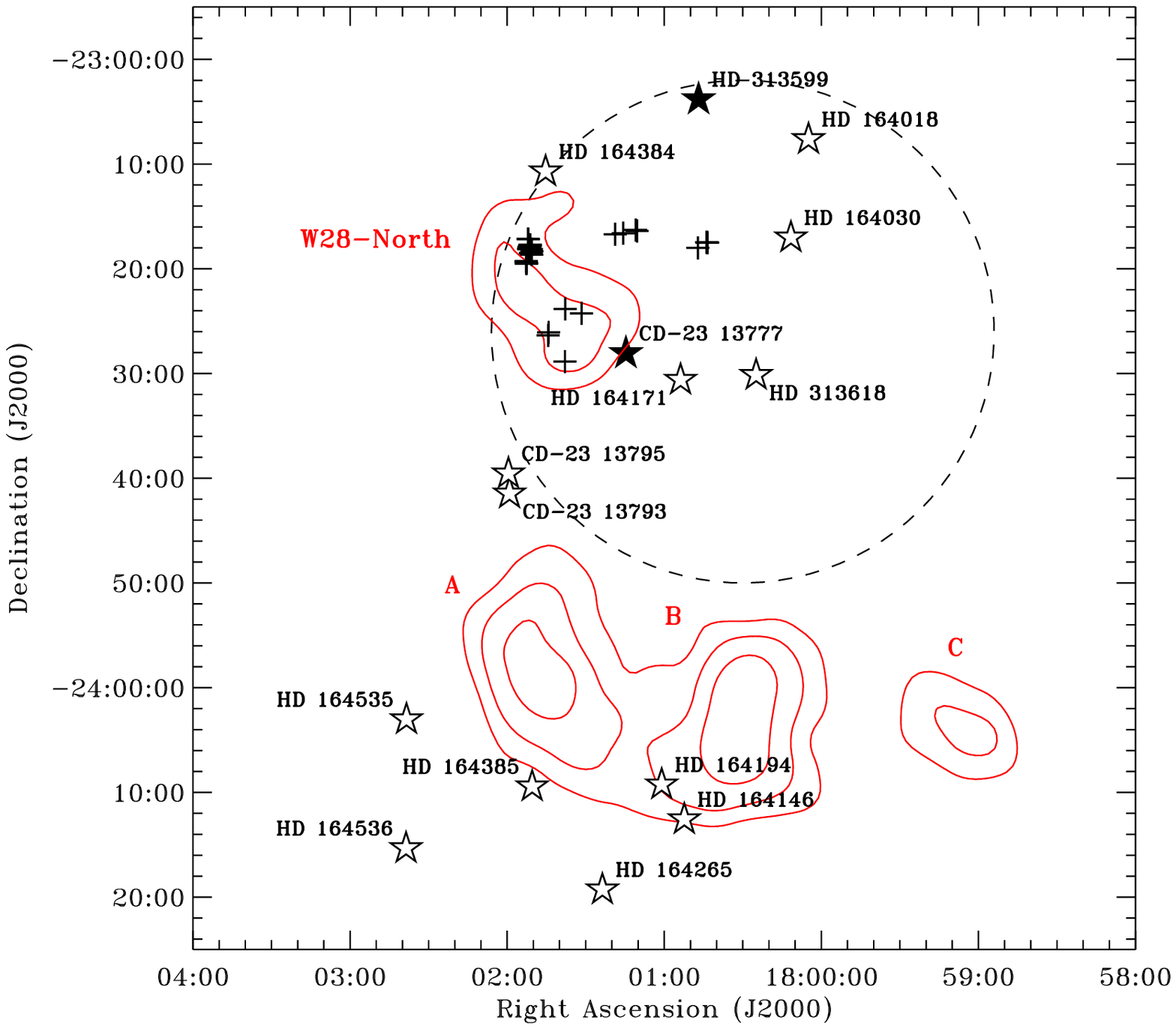}
\caption{On-sky locations of the 15 early-type stars in the W28 region surveyed with ARCES. Solid symbols are used for stars whose \emph{Gaia} DR2 parallaxes place them at distances greater than 1.9~kpc. Stars represented by open symbols are likely foreground targets. Red contours indicate the 4, 5, and 6$\sigma$ significance levels of the TeV sources observed by HESS (Aharonian et al.~2008) with labels as in Figure~\ref{fig:dss_hess}. Plus signs indicate the positions of the 41 1720 MHz OH masers observed by Claussen et al.~(1997). The dashed circle gives the approximate position and size of the radio shell associated with W28 (Green 2019).}
\label{fig:w28_targets}
\end{figure}

Fifteen stars with on-sky positions near W28 were observed with ARCES over the course of several years, mainly between 2011 September and 2013 April. (Additional observations were obtained of CD$-$23~13777 in 2015 May and August.) Details concerning the target stars and the ARCES observations are presented in Table~\ref{tab:arces_targets}. The on-sky locations of the targets in relation to the radio shell of W28, and the TeV sources observed by HESS, are shown in Figure~\ref{fig:w28_targets}. The original concept was to survey stars that could serve as probes of the kinematics and physical conditions in the clouds interacting with W28. Thus, targets were chosen that were early-type stars that were bright enough for ARCES observations (i.e., $V\lesssim11$) and had spectroscopic distances that placed them either within or behind the SNR (i.e., $d\gtrsim1.9$~kpc). We deliberately chose targets that could potentially probe both the northern radio shell of W28, including the northern TeV source (W28-North), and the southern HESS sources (A, B, and C).

\begin{figure*}
\includegraphics[width=\textwidth]{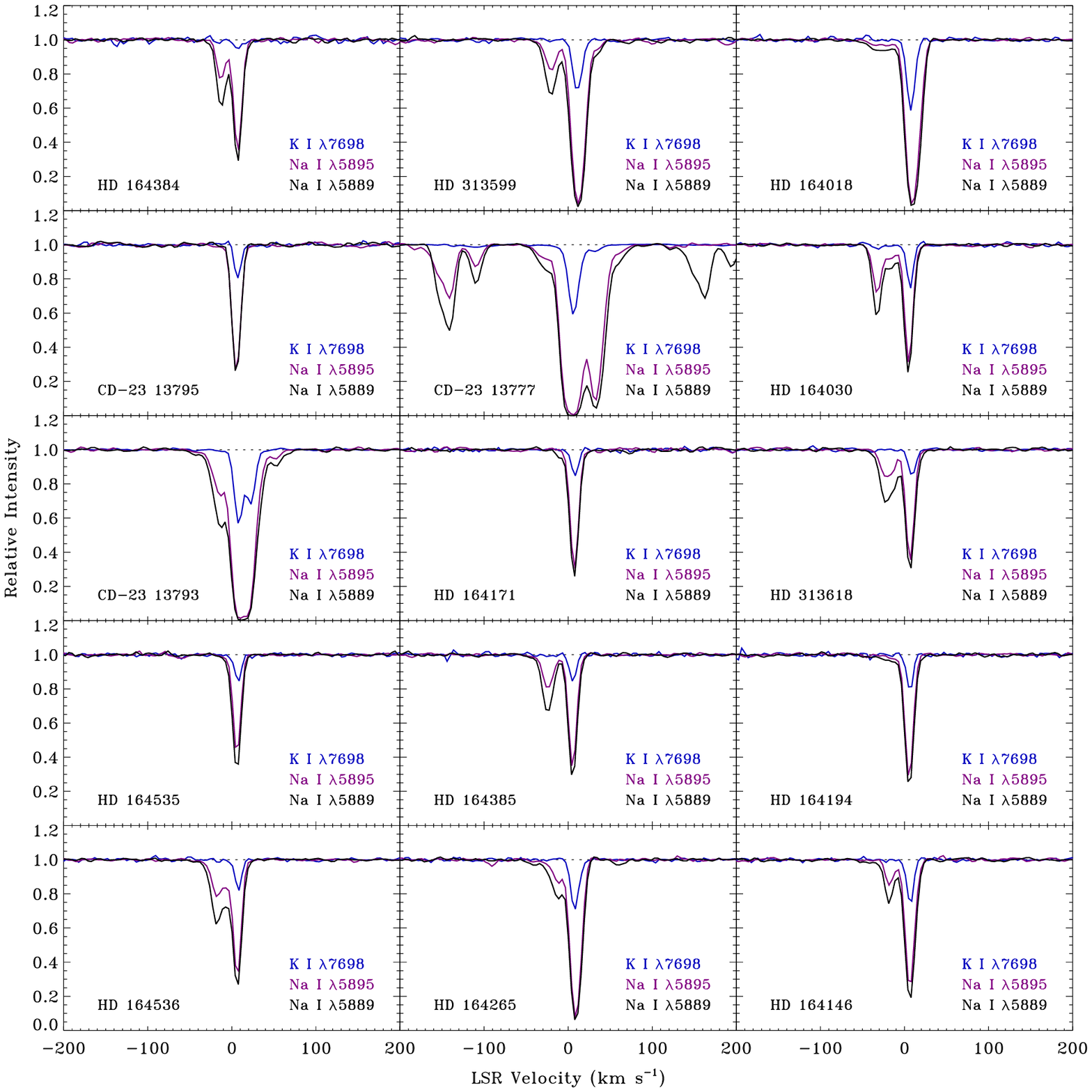}
\caption{ARCES spectra of the Na~{\sc i}~$\lambda\lambda5889,5895$ and K~{\sc i}~$\lambda7698$ lines toward 15 stars in the W28 region. The panels are arranged to approximate the on-sky positions of the targets as shown in Figure~\ref{fig:w28_targets}. Only toward CD$-$23~13777 do we find evidence of high-velocity interstellar absorption. (Note the overlap between the high negative velocity $\lambda5895$ features and the high positive velocity $\lambda5889$ components in this direction.) The weak absorption features outside the velocity range of the main interstellar components toward HD~164265 are weak stellar absorption lines.}
\label{fig:w28_spectra}
\end{figure*}

The raw echelle observations were reduced following the same procedures as described in Section~\ref{sec:observations}. The final, reduced ARCES spectra of the Na~{\sc i}~$\lambda\lambda5889,5895$ and K~{\sc i}~$\lambda7698$ lines toward all 15 targets are displayed in Figure~\ref{fig:w28_spectra}. As is readily apparent from the figure, high-velocity (>100~km~s$^{-1}$) interstellar absorption is detected only toward CD$-$23~13777. (The same conclusion is reached if one examines the Ca~{\sc ii} absorption lines.) Initially, this finding was somewhat surprising. In other cases where a SNR is interacting with diffuse atomic and/or molecular clouds in its vicinity, one can typically find multiple sight lines displaying high-velocity Na~{\sc i} and Ca~{\sc ii} absorption features (e.g., near the Vela SNR and IC~443; Danks \& Sembach 1995; Cha \& Sembach 2000; Welsh \& Sallmen 2003; Hirschauer et al.~2009; Ritchey et al.~2020). However, after reliable distances to our target stars became available, from trigonometric parallaxes measured by the \emph{Gaia} satellite (see Table~\ref{tab:arces_targets}), it became clear that most of the stars we observed were foreground stars. (Hence, the majority of this paper focuses on the line of sight toward CD$-$23~13777.) For completeness, we present in Table~\ref{tab:arces_ew} the total equivalent widths of the prominent interstellar absorption lines observed toward all 15 stars surveyed with ARCES.

\begin{table*}
\centering
\caption{Stellar and observational data for the 15 targets in the W28 region surveyed with ARCES. Spectral types and $B$ and $V$ magnitudes are from the SIMBAD database. Values of $E$($B$$-$$V$) were calculated using intrinsic colors from Wegner (1994). Since the luminosity class of HD 313618 is not known, the reddening value is only approximate. The distances listed are from trigonometric parallaxes measured by the \emph{Gaia} satellite (Bailer-Jones et al.~2018). The last column gives the total exposure time on each target for the ARCES observations.}
\label{tab:arces_targets}
\begin{tabular}{lcccccccc}
\hline
Star & Sp.~Type & R.A. & Dec. & $B$ & $V$ & $E$($B$$-$$V$) & $d$ & Exp.~Time \\
 & & (J2000) & (J2000) & (mag) & (mag) & (mag) & (kpc) & (s) \\
\hline
HD 164018 & B1/2 Ib & 18 00 04.91 & $-$23 07 36.9 & 9.86 & 9.26 & 0.77 & $1.3_{-0.1}^{+0.1}$ & 1200 \\
HD 164030 & B5 III & 18 00 11.62 & $-$23 17 00.2 & 9.74 & 9.72 & 0.17 & $1.1_{-0.1}^{+0.1}$ & 2100 \\
HD 164146 & B2 II/III & 18 00 52.34 & $-$24 12 33.6 & 8.15 & 8.19 & 0.14 & $0.10_{-0.01}^{+0.01}$ & \phantom{1}600 \\
HD 164171 & B8/9 II & 18 00 53.76 & $-$23 30 35.8 & 9.90 & 9.85 & 0.07 & $0.66_{-0.02}^{+0.02}$ & 1800 \\
HD 164194 & B3 II/III & 18 01 00.97 & $-$24 09 14.3 & 8.69 & 8.62 & 0.22 & $1.7_{-0.2}^{+0.3}$ & \phantom{1}600 \\
HD 164265 & B8/9 Ib/II & 18 01 23.61 & $-$24 19 15.8 & 9.66 & 9.29 & 0.39 & $1.5_{-0.1}^{+0.2}$ & 1200 \\
HD 164384 & B1/2 Ib/II & 18 01 45.32 & $-$23 10 42.0 & 8.19 & 8.26 & 0.10 & $1.1_{-0.1}^{+0.1}$ & \phantom{1}600 \\
HD 164385 & B2 III & 18 01 50.44 & $-$24 09 25.3 & 8.04 & 8.07 & 0.16 & $1.7_{-0.2}^{+0.2}$ & \phantom{1}600 \\
HD 164535 & B5/7 Iab/II & 18 02 38.45 & $-$24 03 00.8 & 9.83 & 9.84 & 0.05 & $1.1_{-0.1}^{+0.1}$ & 2100 \\
HD 164536 & O7.5 V(n)z & 18 02 38.63 & $-$24 15 19.4 & 7.37 & 7.40 & 0.26 & $1.4_{-0.5}^{+1.5}$ & \phantom{1}300 \\
HD 313599 & B0.5 Ia & 18 00 46.89 & $-$23 03 48.3 & 10.51\phantom{1} & 9.69 & 1.02 & $2.2_{-0.3}^{+0.3}$ & 1800 \\
HD 313618 & B5 & 18 00 24.90 & $-$23 30 08.2 & 9.62 & 9.64 & $\sim$0.1 & $0.92_{-0.05}^{+0.05}$ & 3600 \\
CD$-$23 13777 & O9 Ib & 18 01 14.63 & $-$23 28 03.0 & 11.64\phantom{1} & 10.55\phantom{1} & 1.36 & $2.4_{-0.3}^{+0.3}$ & 23400\phantom{1} \\
CD$-$23 13793 & O6 III & 18 01 59.09 & $-$23 41 27.7 & 11.58\phantom{1} & 10.32\phantom{1} & 1.56 & $1.1_{-0.5}^{+3.0}$ & 5400 \\
CD$-$23 13795 & A2 & 18 01 59.51 & $-$23 39 32.7 & \ldots & 11.3\phantom{11} & \ldots & $0.45_{-0.01}^{+0.01}$ & 7200 \\
\hline
\end{tabular}
\end{table*}

\begin{table*}
\centering
\caption{Total equivalent widths (in m\AA{}) of the Na~{\sc i}~$\lambda\lambda5889,5895$, Ca~{\sc ii}~$\lambda\lambda3933,3968$, K~{\sc i}~$\lambda7698$, Ca~{\sc i}~$\lambda4226$, CH~$\lambda4300$, and CH$^+$~$\lambda4232$ lines toward the 15 stars in the W28 region surveyed with ARCES. Errors in equivalent width are calculated as the product of the width of each individual component and the root mean square of the noise in the continuum. The individual component errors are then added in quadrature to yield the total equivalent width errors. 3$\sigma$ upper limits are given in cases where a line is not detected.}
\label{tab:arces_ew}
\begin{tabular}{lcccccccc}
\hline
Star & $W_{\lambda}$(5889) & $W_{\lambda}$(5895) & $W_{\lambda}$(3933) & $W_{\lambda}$(3968) & $W_{\lambda}$(7698) & $W_{\lambda}$(4226) & $W_{\lambda}$(4300) & $W_{\lambda}$(4232) \\
\hline
HD 164018 & $466.5\pm2.6$ & $397.7\pm2.1$ & $243.9\pm7.6$ & $167.9\pm8.0$ & $125.9\pm2.5$\phantom{1} & $\lesssim8.0$ & $22.8\pm2.3$\phantom{1} & $40.3\pm4.2$\phantom{1} \\
HD 164030 & $344.3\pm1.9$ & $260.2\pm1.9$ & $309.2\pm4.2$ & $185.2\pm4.5$ & $67.8\pm2.5$ & $4.7\pm1.0$ & $8.9\pm2.0$ & $11.0\pm0.8$\phantom{1} \\
HD 164146 & $316.1\pm2.4$ & $239.3\pm2.1$ & $246.1\pm4.0$ & $151.3\pm3.4$ & $66.2\pm1.6$ & $5.6\pm1.2$ & $8.4\pm1.2$ & $10.0\pm1.0$\phantom{1} \\
HD 164171 & $194.1\pm1.6$ & $168.9\pm1.5$ & $100.4\pm1.9$ & \phantom{1}$66.8\pm2.8$ & $40.0\pm1.8$ & $\lesssim4.0$ & $6.0\pm1.3$ & $9.0\pm0.9$ \\
HD 164194 & $231.1\pm6.0$ & $183.2\pm2.7$ & $169.4\pm4.1$ & \phantom{1}$92.3\pm4.0$ & $50.7\pm1.9$ & $4.6\pm1.1$ & $7.1\pm1.3$ & $8.7\pm1.3$ \\
HD 164265 & $409.1\pm2.6$ & $336.5\pm2.3$ & $378.1\pm5.9$ & $240.6\pm8.6$ & $89.2\pm1.9$ & $8.2\pm2.3$ & $18.4\pm2.4$\phantom{1} & $23.6\pm2.1$\phantom{1} \\
HD 164384 & $289.0\pm2.5$ & $216.8\pm2.4$ & $172.3\pm3.5$ & \phantom{1}$98.1\pm3.6$ & $17.7\pm4.0$ & $\lesssim5.2$ & $\lesssim5.5$ & $\lesssim4.8$ \\
HD 164385 & $296.2\pm2.3$ & $213.9\pm2.0$ & $238.6\pm4.1$ & $141.7\pm3.0$ & $39.7\pm1.9$ & $3.2\pm0.8$ & $4.7\pm0.8$ & $3.3\pm1.1$ \\
HD 164535 & $168.7\pm1.4$ & $129.5\pm1.2$ & $112.6\pm2.7$ & \phantom{1}$77.1\pm2.4$ & $39.2\pm1.5$ & $2.6\pm0.8$ & $6.1\pm1.7$ & $11.6\pm1.4$\phantom{1} \\
HD 164536 & $346.4\pm2.1$ & $247.8\pm1.9$ & $266.5\pm4.5$ & $154.5\pm4.2$ & $43.3\pm1.4$ & $4.6\pm1.0$ & $3.6\pm0.7$ & $10.4\pm1.3$\phantom{1} \\
HD 313599 & $555.0\pm4.3$ & $448.5\pm4.0$ & $398.5\pm8.0$ & $271.7\pm7.3$ & $103.8\pm2.0$\phantom{1} & $36.4\pm5.6$\phantom{1} & $24.9\pm2.7$\phantom{1} & $47.9\pm3.5$\phantom{1} \\
HD 313618 & $320.4\pm2.5$ & $220.2\pm2.3$ & $272.7\pm5.7$ & $174.4\pm5.2$ & $42.5\pm1.7$ & $\lesssim6.5$ & $4.0\pm1.1$ & $10.5\pm1.8$\phantom{1} \\
CD$-$23 13777 & $1486.7\pm3.4$\phantom{1} & $1205.4\pm3.4$\phantom{1} & $1034.1\pm12.8$ & \phantom{1}$692.5\pm10.5$ & $182.8\pm1.6$\phantom{1} & $47.6\pm4.9$\phantom{1} & $54.0\pm3.4$\phantom{1} & $139.5\pm4.1$\phantom{1}\phantom{1} \\
CD$-$23 13793 & $885.6\pm5.6$ & $742.3\pm5.4$ & \phantom{1}$525.0\pm18.7$ & \phantom{1}$378.4\pm12.8$ & $243.6\pm1.8$\phantom{1} & $14.8\pm2.2$\phantom{1} & $49.2\pm3.8$\phantom{1} & $87.7\pm3.6$\phantom{1} \\
CD$-$23 13795 & $191.2\pm1.7$ & $189.5\pm1.7$ & \phantom{1}$71.7\pm5.4$ & \phantom{1}$46.1\pm3.6$ & $48.6\pm1.5$ & $\lesssim5.9$ & $5.8\pm1.7$ & $12.2\pm1.4$\phantom{1} \\
\hline
\end{tabular}
\end{table*}

The star HD~313599 has a \emph{Gaia} DR2 distance of $2.2\pm0.3$~kpc and so may also lie behind the SNR (whose distance is estimated to be $1.9\pm0.3$~kpc; Vel\'{a}zquez et al.~2002). This star lies along the northern edge of the radio shell associated with W28 (Figure~\ref{fig:w28_targets}). Thus, any shocked gas in this direction may have a large transverse component to its motion and would therefore not appear to be at high velocity from our perspective. Moreover, the shocked gas components would be blended with absorption from quiescent gas near the systemic velocity of W28. There is a weak Na~{\sc i} (and Ca~{\sc ii}) absorption component near +35~km~s$^{-1}$ toward HD~313599 that is not seen toward most of the other stars in the region (Figure~\ref{fig:w28_spectra}). This component could trace either a low-velocity shock or a high-velocity shock with a large transverse component. The low $N$(Na~{\sc i})/$N$(Ca~{\sc ii}) ratio for this component ($\sim$0.2) supports this hypothesis.

The star CD$-$23~13793 lies along the southeast boundary of the radio shell associated with W28 (Figure~\ref{fig:w28_targets}). While this sight line exhibits none of the very high velocity Na~{\sc i} and Ca~{\sc ii} absorption components seen toward CD$-$23~13777, it does display fairly strong and complex absorption extending from $-$60 to +70~km~s$^{-1}$ (Figure~\ref{fig:w28_spectra}). The distance to this star derived from the \emph{Gaia} DR2 parallax is $1.1_{-0.5}^{+3.0}$~kpc, which would suggest that the star lies in front of the SNR. However, the \emph{Gaia} distance has a large associated uncertainty (with an upper bound equal to $\sim$4~kpc). Given the strong and complex interstellar absorption seen in this direction, and the high degree of interstellar reddening (see Table~\ref{tab:arces_targets}), it is probably more likely that this star does in fact lie behind the SNR (and the associated molecular clouds with which it is interacting).

\section{Detailed Component Structure toward CD$-$23~13777}\label{sec:appendix_b}
In Table~\ref{tab:comp_struct}, we present the component parameters derived from profile synthesis fits to the interstellar absorption lines detected toward CD$-$23~13777. In particular, we list the LSR velocity, the column density, and the Doppler $b$-value for each of the components included in the fits for the different species. The fits themselves are presented in Figures~\ref{fig:profile_fits1} and \ref{fig:profile_fits2}. We do not list column densities for the main Na~{\sc i} components at low velocity because the absorption associated with these components is too strongly saturated to allow for accurate column density determinations. Furthermore, we caution that the low resolution of the ARCES spectra makes it difficult to derive a unique set of solutions for the component structures of the complex blended absorption lines seen toward CD$-$23~13777. Our analysis of these features was guided by expectations concerning the typical Doppler widths found from high-resolution studies of interstellar absorption lines (e.g., Welty \& Hobbs 2001) and we have aimed for consistency among the species observed at similar velocities. Still, we acknowledge that the detailed underlying component structure of the absorption lines detected toward CD$-$23~13777 may not be known precisely. This is why in Section~\ref{subsec:col_den} we focus only on the total column densities of those nine distinct absorption features identified along the line of sight (Table~\ref{tab:components}). While the fractional column densities of the individual subcomponents that constitute each distinct feature are only poorly constrained, the total column densities of those absorption features are much more robust. Higher resolution optical spectra of CD$-$23~13777 would be needed to confirm (and improve upon) the component structures presented in Table~\ref{tab:comp_struct}.

\begin{landscape}
\begin{table}
\centering
\caption{Velocities (in km~s$^{-1}$), column densities (in cm$^{-2}$), and Doppler $b$-values (in km~s$^{-1}$) derived from profile synthesis fits to the interstellar absorption lines detected toward CD$-$23~13777. The Ca~{\sc ii} and Na~{\sc i} column densities listed here were obtained by taking the weighted means of the results from the two lines of the doublets. (No column densities are given for the main Na~{\sc i} components between $-$16 and +41~km~s$^{-1}$ because the absorption associated with these components is too strongly saturated to allow for accurate column density determinations.)}
\label{tab:comp_struct}
\begin{tabular}{cccccccccccccccccc}
\hline
$v_{\rm LSR}$ & $\log N$(Ca~{\sc ii}) & $b$ & $v_{\rm LSR}$ & $\log N$(Na~{\sc i}) & $b$ & $v_{\rm LSR}$ & $\log N$(Ca~{\sc i}) & $b$ & $v_{\rm LSR}$ & $\log N$(K~{\sc i}) & $b$ & $v_{\rm LSR}$ & $\log N$(CH$^+$) & $b$ & $v_{\rm LSR}$ & $\log N$(CH) & $b$ \\
\hline
$-$157.8 & $11.87\pm0.03$ & 4.5 & $-$156.9 & $11.52\pm0.01$ & 2.5 &  \ldots &      \ldots & \ldots &  \ldots &      \ldots & \ldots & \ldots &      \ldots & \ldots & \ldots &      \ldots & \ldots \\
$-$148.5 & $12.21\pm0.03$ & 2.9 & $-$147.2 & $11.75\pm0.01$ & 4.5 &  \ldots &      \ldots & \ldots &  \ldots &      \ldots & \ldots & \ldots &      \ldots & \ldots & \ldots &      \ldots & \ldots \\
$-$137.4 & $12.09\pm0.03$ & 3.6 & $-$138.6 & $11.87\pm0.02$ & 2.9 &  \ldots &      \ldots & \ldots &  \ldots &      \ldots & \ldots & \ldots &      \ldots & \ldots & \ldots &      \ldots & \ldots \\
$-$118.4 & $11.53\pm0.05$ & 3.0 & $-$118.6 & $10.77\pm0.03$ & 3.9 &  \ldots &      \ldots & \ldots &  \ldots &      \ldots & \ldots & \ldots &      \ldots & \ldots & \ldots &      \ldots & \ldots \\
$-$110.9 & $12.23\pm0.03$ & 2.5 & $-$110.6 & $11.49\pm0.01$ & 3.7 &  \ldots &      \ldots & \ldots &  \ldots &      \ldots & \ldots & \ldots &      \ldots & \ldots & \ldots &      \ldots & \ldots \\
$-$102.6 & $11.80\pm0.04$ & 2.1 & $-$103.0 & $10.78\pm0.03$ & 0.5 &  \ldots &      \ldots & \ldots &  \ldots &      \ldots & \ldots & \ldots &      \ldots & \ldots & \ldots &      \ldots & \ldots \\
 \phantom{1}$-$51.3 & $11.12\pm0.10$ & 1.5 &   \ldots &      \ldots & \ldots &  \ldots &      \ldots & \ldots &  \ldots &      \ldots & \ldots & \ldots &      \ldots & \ldots & \ldots &      \ldots & \ldots \\
 \phantom{1}$-$34.0 & $11.70\pm0.04$ & 3.7 &  \phantom{1}$-$35.4 & $11.00\pm0.02$ & 2.8 & $-$32.5 & $10.05\pm0.20$ & 4.5 &  \ldots &      \ldots & \ldots & \ldots &      \ldots & \ldots & \ldots &      \ldots & \ldots \\
 \phantom{1}$-$24.9 & $11.93\pm0.04$ & 1.7 &  \phantom{1}$-$26.3 & $11.13\pm0.02$ & 0.6 & $-$23.4 & $10.06\pm0.17$ & 0.5 &  \ldots &      \ldots & \ldots & \ldots &      \ldots & \ldots & \ldots &      \ldots & \ldots \\
 \phantom{1}$-$14.6 & $11.92\pm0.03$ & 2.4 &  \phantom{1}$-$16.0 &      [sat.] & \ldots & $-$13.1 & $10.22\pm0.12$ & 1.7 & $-$18.8 &  \phantom{1}$9.98\pm0.15$ & 2.2 & \ldots &      \ldots & \ldots & \ldots &      \ldots & \ldots \\
  \phantom{11}$-$2.5 & $12.31\pm0.03$ & 3.7 &   \phantom{11}$-$4.0 &      [sat.] & \ldots &  \phantom{1}$-$1.1 & $10.41\pm0.08$ & 1.2 &  \phantom{1}$-$7.0 & $10.70\pm0.04$ & 4.1 & \phantom{1}$-$3.4 & $13.51\pm0.03$ & 4.5 & \phantom{1}$-$0.9 & $13.27\pm0.04$ & 2.6 \\
    \phantom{11}+4.4 & $12.34\pm0.05$ & 2.5 &     \phantom{11}+3.0 &      [sat.] & \ldots &    \phantom{1}+5.9 & $10.37\pm0.09$ & 1.7 &    \phantom{1}+5.3 & $11.98\pm0.03$ & 3.7 &   \phantom{1}+5.0 & $13.76\pm0.03$ & 3.4 &   \phantom{1}+6.1 & $13.42\pm0.03$ & 3.3 \\
    \phantom{11}+9.1 & $12.41\pm0.04$ & 2.8 &     \phantom{11}+7.7 &      [sat.] & \ldots &   +10.6 & $10.50\pm0.07$ & 1.3 &    \phantom{1}+9.2 & $11.03\pm0.02$ & 1.2 &   \phantom{1}+9.4 & $13.48\pm0.03$ & 4.5 &  +10.8 & $12.56\pm0.19$ & 4.5 \\
   \phantom{1}+14.4 & $12.01\pm0.04$ & 2.3 &    \phantom{1}+13.0 &      [sat.] & \ldots &   +15.9 &  \phantom{1}$9.48\pm0.43$ & 1.0 &   +14.4 & $11.18\pm0.02$ & 1.0 & \ldots &      \ldots & \ldots & \ldots &      \ldots & \ldots \\
   \phantom{1}+19.1 & $12.08\pm0.04$ & 1.9 &    \phantom{1}+17.7 &      [sat.] & \ldots &   +20.6 & $10.25\pm0.11$ & 1.7 &  \ldots &      \ldots & \ldots & \ldots &      \ldots & \ldots & \ldots &      \ldots & \ldots \\
   \phantom{1}+25.7 & $12.07\pm0.03$ & 2.4 &    \phantom{1}+24.3 &      [sat.] & \ldots &   +27.2 &  \phantom{1}$8.97\pm0.80$ & 0.5 &   +24.4 & $10.43\pm0.06$ & 0.5 &  +28.0 & $13.18\pm0.06$ & 3.4 &  +27.4 & $12.90\pm0.10$ & 4.5 \\
   \phantom{1}+32.6 & $12.15\pm0.03$ & 2.9 &    \phantom{1}+31.2 &      [sat.] & \ldots &   +34.1 & $10.45\pm0.09$ & 4.5 &   +32.9 & $10.87\pm0.03$ & 4.5 &  +33.5 & $13.56\pm0.04$ & 1.4 &  +34.3 & $13.13\pm0.06$ & 4.5 \\
   \phantom{1}+42.8 & $11.70\pm0.04$ & 2.3 &    \phantom{1}+41.3 &      [sat.] & \ldots &  \ldots &      \ldots & \ldots &  \ldots &      \ldots & \ldots &  +38.1 & $13.28\pm0.05$ & 1.0 & \ldots &      \ldots & \ldots \\
   \phantom{1}+55.4 & $10.85\pm0.17$ & 2.5 &    \phantom{1}+56.2 & $11.22\pm0.01$ & 4.5 &   +57.4 &  \phantom{1}$9.96\pm0.20$ & 1.7 &  \ldots &      \ldots & \ldots &  +60.4 & $12.81\pm0.11$ & 2.0 & \ldots &      \ldots & \ldots \\
   \phantom{1}+66.1 & $10.72\pm0.21$ & 1.9 &    \phantom{1}+66.9 & $10.46\pm0.04$ & 0.5 &  \ldots &      \ldots & \ldots &  \ldots &      \ldots & \ldots & \ldots &      \ldots & \ldots & \ldots &      \ldots & \ldots \\
  +119.3 & $11.78\pm0.04$ & 1.2 &   +121.1 & $10.62\pm0.03$ & 0.5 &  \ldots &      \ldots & \ldots &  \ldots &      \ldots & \ldots & \ldots &      \ldots & \ldots & \ldots &      \ldots & \ldots \\
  +136.6 & $11.69\pm0.05$ & 1.2 &   +136.4 & $10.44\pm0.05$ & 0.5 &  \ldots &      \ldots & \ldots &  \ldots &      \ldots & \ldots & \ldots &      \ldots & \ldots & \ldots &      \ldots & \ldots \\
\hline
\end{tabular}
\end{table}
\end{landscape}

%%%%%%%%%%%%%%%%%%%%%%%%%%%%%%%%%%%%%%%%%%%%%%%%%%

% Don't change these lines
\bsp	% typesetting comment
\label{lastpage}
\end{document}